\RequirePackage{fix-cm}
\documentclass[smallextended]{svjour3}       
\smartqed  
\usepackage{graphicx}
\usepackage{SIunits}
\journalname{Optical and Quantum Electronic}
\begin{document}

\title{Robust design of Si/Si$_3$N$_4$ high contrast grating mirror for mid-infrared VCSEL application
}

\titlerunning{Robust design of Si/Si$_3$N$_4$ HCG mirror for mid-IR VCSEL application}        

\author{Chevallier C. \and Fressengeas N. \and Genty~F. \and Jacquet J.
%
}                     

\institute{Chevallier C. \and Genty F. \and Jacquet J. \at
              Sup\'elec, 2 rue Edouard Belin 57070 Metz, France\\
              \email{christyves.chevallier@supelec.fr}
           \and Fressengeas N. \at
              LMOPS, Laboratoire Mat\'eriaux Optiques, Photonique et Syst\`emes, EA 4423, Unit\'e de Recherche Commune \`a l'Universit\'e Paul Verlaine - Metz et Sup\'elec, 2 rue Edouard Belin 57070 Metz, France
}

\date{Received: date / Accepted: date}

\maketitle

\begin{abstract}
  A Si/Si$_3$N$_4$ high contrast grating mirror has been designed for a VCSEL integration
  in mid-infrared ($\lambda = $\unit{2.65}{\micro\meter}).
  The use of an optimization algorithm which maximizes a VCSEL mirror quality factor
  allowed the adjustment of the grating parameters while keeping large and shallow
  grating pattern. The robustness with respect to fabrication error has been enhanced
  thanks to a precise study of the grating dimension tolerances. The final mirror exhibits
  large high reflectivity bandwidth with a polarization selectivity and several percent
  of tolerance on the grating dimensions.

\end{abstract}

\section{Introduction}
  \label{intro}

  High contrast grating (HCG) mirrors with a grating period smaller or in the range of the reflected wavelength
  can be designed to exhibit very high performances
  with energy only in the 0-order of diffraction. HCGs which are made of a high index grating 
  on top of a low index sublayer can achieve a
  reflectivity as high as 99.9~\% and bandwidths larger than 100~nm
  together with a strong polarization selectivity between TM and TE mode~\cite{mateus_ptl_2004,chevallier_apa_2010}.
  These characteristics are very interesting for replacing the top Bragg mirror in VCSEL devices~\cite{huang_np_2007}.
  Moreover, the circular VCSEL structure
  usually presents no polarisation selectivity and the use of
  a grating should improve the stability of the emitted beam~\cite{debernardi_jstqe_2005}.
  Also, the wavelength of emission of mid-IR electrically-pumped VCSEL 
  is currently limited close to \unit{2.6}{\micro\meter}
  since the necessary semiconductor Bragg mirrors become as thick as \unit{11}{\micro\meter}
  with more than 20 quarter wavelength pairs~\cite{ducanchez_el_2009_2,bachmann_njp_2009} making the epitaxial growth control critical.
  With a reduction of a factor of 5 of the thickness of the top mirror and 
  fabrication tolerances as large as 20~\%~\cite{zhou_ptl_2008,chevallier_olt_2011}, the use of HCGs 
  should allow VCSEL operation at larger wavelengths. Moreover, the HCG dimensions are scalable with the wavelength 
  in the limit of the refractive index dispersion~\cite{mateus_ptl_2004} which makes the designs presented in this work scalable for other wavelengths.

  In this work, we design a Si/Si$_3$N$_4$ HCG for a VCSEL application at \unit{2.65}{\micro\meter}.
  First, the optical properties that VCSEL mirrors must exhibit 
  are defined and quantitatively expressed through a quality factor. 
  The dimensions of the HCG structure are then adjusted by an optimization algorithm 
  to maximize the quality factor and meet VCSEL requirements. Finally, the HCG robustness is enhanced
  through a precise study of the tolerance on the physical dimensions.
  The final design has thus an optimized reflectivity with several percent 
  of tolerance as an error margin that the technological manufacturing process must respect.

\section{Mirror design}
\begin{figure}[!t]
    \centering
    \includegraphics[width=1.3in]{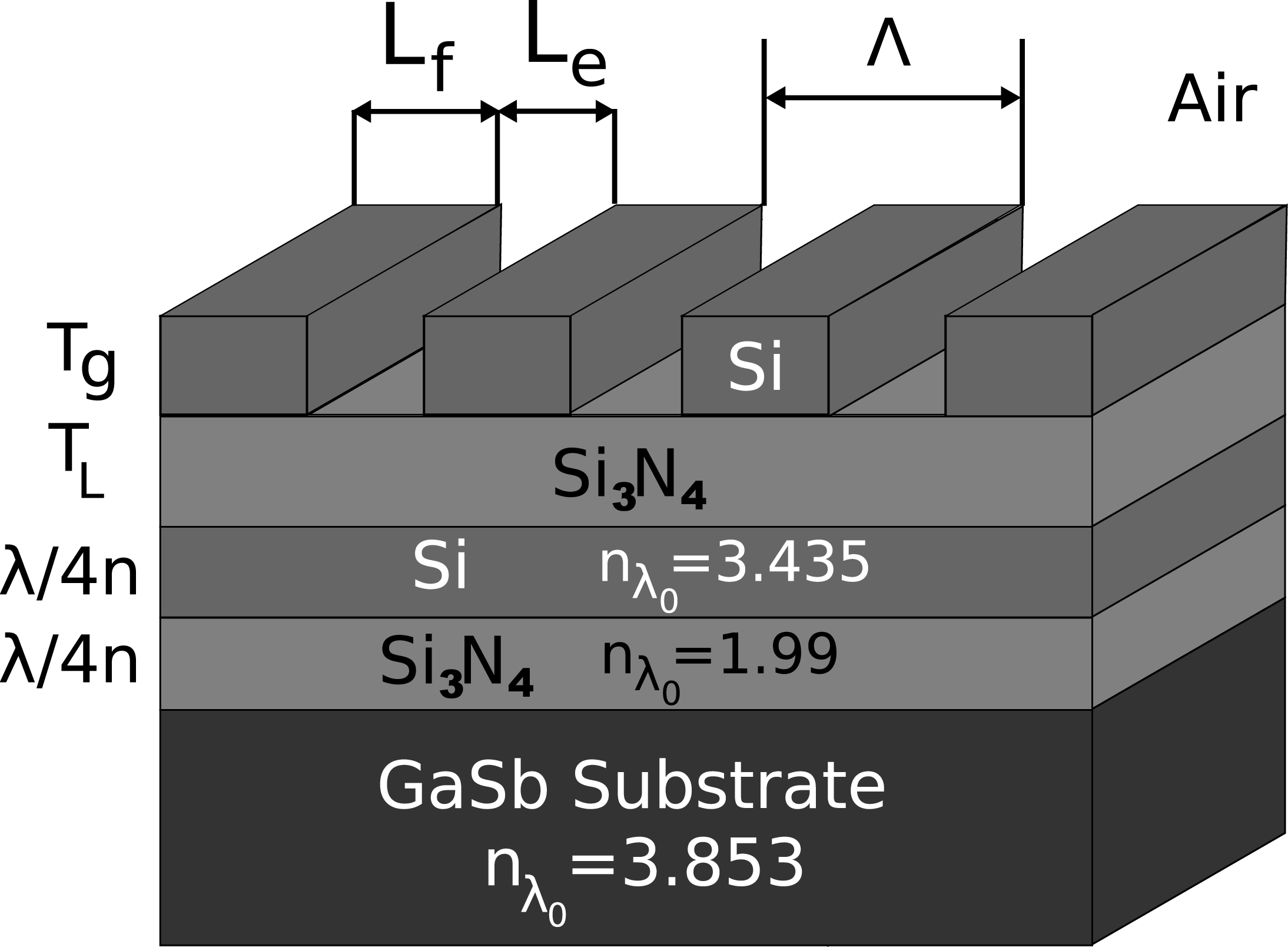}
    \caption{Scheme of the HCG optimized for $\lambda_0 =$~\unit{2.65}{\micro\meter} exhibiting a 218~nm large bandwidth
	with $L_e =$~\unit{829}{\nano\meter}, $L_f =$~\unit{522}{\nano\meter}, $T_g =$~\unit{899}{\nano\meter}, $T_L =$~\unit{896}{\nano\meter}.
	}
    \label{fig_scheme}
  \end{figure}

  In a previous works~\cite{chevallier_olt_2011}, Si/SiO$_2$ structures were designed 
  for applications at $\lambda = $\unit{2.65}{\micro\meter}. 
  However in the \unit{2.65-3}{\micro\meter} wavelength range, OH-radicals present an absorption band~\cite{soref_joa_2006}
  and could decrease the HCG performances depending on their concentration in the SiO$_2$ layer~\cite{chevallier_olt_2011}.
  The Si$_3$N$_4$ should be well suitable for the low index material since it exhibits no absorption at these wavelengths~\cite{soref_joa_2006}.
  The mirror structure presented in Fig. \ref{fig_scheme} is made of a silicium grating ($n = 3.435$) on top of a Si$_3$N$_4$ layer
  as the low index sublayer ($n = 1.99$). A Si/Si$_3$N$_4$ quarter wavelength pair has been added below the HCG
  in order to increase the width of the reflectivity bandwidth~\cite{chevallier_apa_2010}. The mirror structure has thus several 
  dimension lengths to define : the etched length $L_e$, the filled length $L_f$, the grating thickness $T_g$ and
  the sublayer thickness $T_L$. Contrary to Bragg mirrors, 
  HCG design rules to achieve high reflectivity are not straightforward and require an optimization process to match the expected properties.

  The VCSEL application requires specific properties on the optical performances of 
  the mirror in order to create a cavity of sufficient quality for laser emission.
  Due to the thin gain region of the VCSEL structure, the reflectivity of the mirror
  is typically chosen to be as high as 99.5~\% for the largest possible bandwidth.
  However, a 99.9~\% reflectivity threshold has been chosen to ensure a fabrication security margin.
  Moreover, due to their one dimensional symmetry, HCG can be shaped to obtain a polarization selectivity.
  Thus, the reflectivity threshold of 99.9~\% is applied only for transverse magnetic (TM)
  mode while keeping a transverse electric (TE) reflectivity below  90~\%, 
  so that the laser effect should be possible only in TM polarization.

  The mirror is simulated by the rigorous coupled wave analysis method (RCWA)~\cite{moharam_josaa_1995,mrcwa} 
  which computes reflection spectra for TM and TE modes for perfectly squared grating with 
  an infinite number of periods. Calculations were done by using constant index values.
  
  Through a numerical analysis of the computed reflection spectra, the quality of a design is quantitatively evaluated 
  by the use of a quality factor Q~\cite{chevallier_apa_2010}:

  \begin{equation}
    \label{eq_q}
    Q = \frac{\Delta\lambda}{\lambda_0}\frac{1}{N}\sum_{\lambda=\lambda_1}^{\lambda_2}{R_{TM}(\lambda)g(\lambda)}
  \end{equation}

  This quality factor is written to take into account the previously defined VCSEL requirements with a 
  polarization selectivity performed by selecting the range of wavelengths $\Delta\lambda$
  around $\lambda_0$ for which $R_{TM}$ is higher than 99.9~\% and $R_{TE}$ is lower than 90~\%. 
  This selection starts at the aimed wavelength $\lambda_0$ and is enlarged both for larger and smaller wavelengths around $\lambda_0$
  while the reflection conditions are met giving two boundaries : $\lambda_1$ and $\lambda_2$.
  The centering of the bandwidth around $\lambda_0 = $ \unit{2.65}{\micro\meter} is made by computing a gaussian weighted average
  of the $R_{TM}$ values on the $N$ points of the bandwidth $\left [ \lambda_1,\lambda_2\right ]$.
  Finally the influence of the width of the stopband is expressed in this quality factor by multiplying the latter average by the bandwidth
  itself $\Delta\lambda~=~|\lambda_2~-~\lambda_1|$ and normalized by $\lambda_0$.

\section{Optimization}

  \begin{table} 
    \caption{Dimensions and tolerances of the optimum mirror.}
    \label{table1}  
    \begin{center}
    
    \begin{tabular}{lllll}
      \hline\noalign{\smallskip}
	Parameter & Optimum & Minimum & Maximum & Tolerance\\
      \noalign{\smallskip}\hline\noalign{\smallskip}
	$L_e$ & 829 nm & 763 nm & 831 nm & $\pm$ 0.2 \%\\
	$L_f$ & 522 nm & 498 nm & 522 nm & $\pm$ 0 \%\\
	$T_g$ & 899 nm & 897 nm & 941 nm & $\pm$ 0.2 \%\\
	$T_L$ & 896 nm & 806 nm & 976 nm & $\pm$ 8.9 \%\\
      \noalign{\smallskip}\hline
	$\Delta\lambda$ & \multicolumn{3}{l}{218 nm}\\
	$\Delta\lambda / \lambda_0$ & \multicolumn{3}{l}{8.3 \%}\\
      \noalign{\smallskip}\hline

      \end{tabular}
    \end{center}
  \end{table}

  The optimization of the mirror performance is made by adjusting the grating parameters to maximize the quality factor $Q$.
  However, with its numerous local maxima, its non linearity and unknown derivative, 
  this Q factor can only be optimized using a global maximization method.
  The Differential Evolution algorithm~\cite{openopt,feoktistov_book_de} has been used as optimization algorithm to find
  automatically the best set of parameters which maximizes the mirror quality.

  The Si/Si$_3$N$_4$-based structure presented in Fig. \ref{fig_scheme} has been optimized under constraints
  by using large and shallow grooves ($L_f , L_e > 500$~nm and an aspect ratio $ T_g / L_e < 1.1$). 
  These limitations are relatively severe for the optimization algorithm since the Si/SiO$_2$ HCGs 
  are typically defined with narrower patterns of less than 260~nm and large aspect ratio of more than 2.6~\cite{mateus_ptl_2004}.
  Since HCGs require a vertical etching profile with squared pattern to reach the 99.9~\% high reflectivity,
  the technological constraints were chosen to obtain an experimental square grating profile as close as possible to the theoretical
  one. However, these values can be easily adjusted before the optimization by the manufacturers according
  to the etching process used and the machine specifications.

  The resulting optimum grating parameters (Table~\ref{table1}) are $L_e =$~\unit{829}{\nano\meter},
  $L_f =$~\unit{522}{\nano\meter}, $T_g =$~\unit{899}{\nano\meter}, $T_L =$~\unit{896}{\nano\meter}.
  This mirror design exhibits a 218~nm large bandwidth with $R_{TM} > 99.9 \%$ together with a high
  polarization selectivity by keeping $R_{TE} < 90 \%$ and meets all previously defined VCSEL requirements.

  In order to evaluate the feasibility of the design, a tolerance study has been performed on the dimensions of the grating.
  The tolerance of one parameter is defined as the allowed variation range of this parameter for which 
  the mirror is performant enough for a VCSEL application with $R_{TM}~>~99.9~\%$ and $R_{TE}~<~90~\%$.
  From a fabrication point of view, the most critical parameters of the grating are the etched length $L_e$ and the filled
  length $L_f$ which result from the etching of the slab. The grating thickness can be controlled with more precision
  during the growth and a selective etching process technique should allow more control on the depth of the etching.

\begin{figure}[!t]
    \centering
    \includegraphics[width=3in]{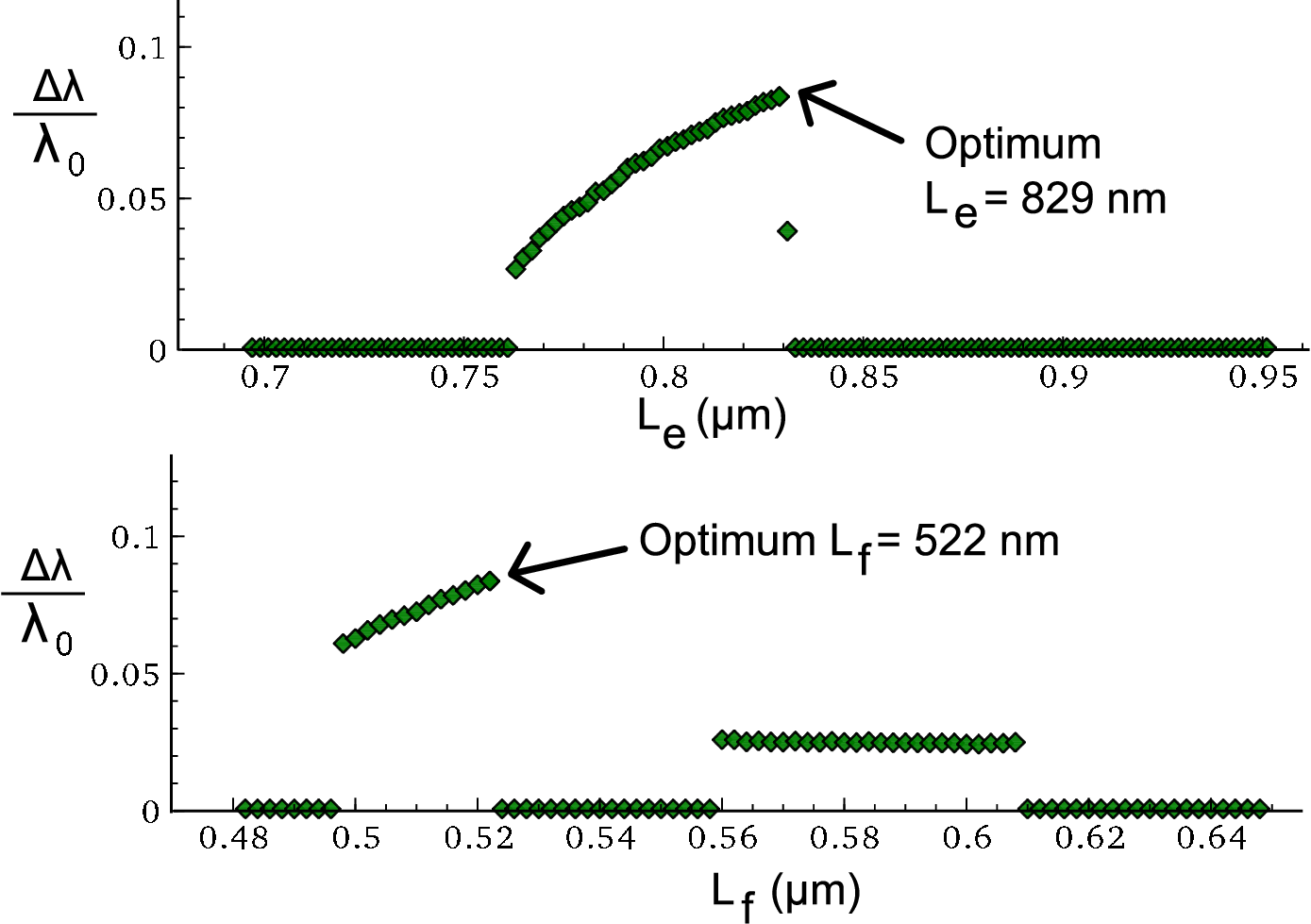}
    \caption{ Evolution of the normalized bandwidth versus the etched length $L_e$
    and the filled length $L_f$. The arrows show the optimum values for
    these parameters which are both at the upper limit of the variation range.
    }
    \label{fig_tolerance1}
  \end{figure}

  The tolerance measurement of the optimized design has shown large variation range of $\Delta L_e = $ 68 nm
  and $\Delta L_f = $ 24 nm. However the optimum point is located at a boundary of these variation ranges and the 
  allowed error is of only 2 nm on $L_e$ and even 0 nm on $L_f$ (Fig. ~\ref{fig_tolerance1}).

\section{Robustness enhancement}

\begin{figure}[!t]
  \centering
  \includegraphics[width=2.5in]{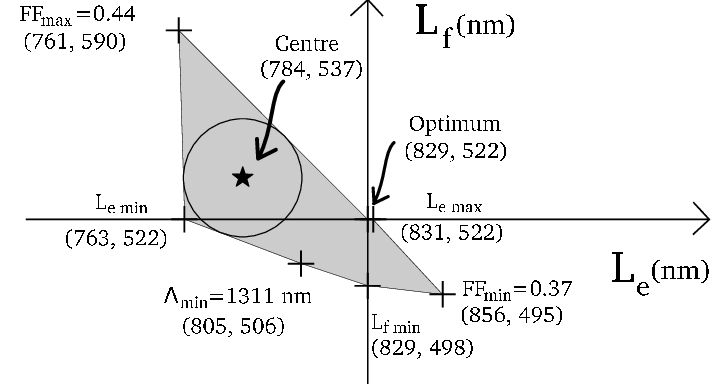}
  \caption{The tolerance map represents the variation range for the $(L_e, L_f)$ couples while keeping $T_g$ and $T_L$ at their optimum value.
      The points (+) define a polygon (in grey) which vertices correspond to the extrema of the variation range of ($L_e$, $L_f$) couples where 
      $R_{TM} $ is below  99.9 \% or $R_{TE}$ above 90 \% at $\lambda_0$ ($\Delta\lambda =$ 0 nm).
      The center of the largest incircle of the polygon ($\star$) indicates the most tolerant point with respect to the fabrication errors on $L_e$ and $L_f$.
  }
  \label{fig_tolerance2}
\end{figure}

\begin{table} 
  \caption{Dimensions and tolerances of the robust mirror.}
  \label{table2}  
  
  \begin{center}
    \begin{tabular}{lllll}
    \hline\noalign{\smallskip}
      Parameter & Value & Tolerance & Minimum & Maximum \\
    \noalign{\smallskip}\hline\noalign{\smallskip}
      $L_e$ & 784 nm & $\pm$ 4 \% & 735 nm & 814 nm\\
      $L_f$ & 537 nm & $\pm$ 5 \% & 512 nm & 598 nm\\
      $T_g$ & 899 nm & $\pm$ 2 \% & 882 nm & 923 nm\\
      $T_L$ & 896 nm & $\pm$ 37 \% & 566 nm & 1496 nm\\
    \noalign{\smallskip}\hline
      $\Delta\lambda$ & \multicolumn{3}{l}{186 nm}\\
      $\Delta\lambda / \lambda_0$ & \multicolumn{3}{l}{7 \%}\\
    \noalign{\smallskip}\hline
    \end{tabular}
  \end{center}
\end{table}

In order to increase the tolerance of the design, a more centered point within the variation 
range should be used. However, during the fabrication process, errors can be made simultaneously
 on several parameters at a time, especially on $L_e$ and $L_f$ if the period was not 
well defined during the photolithography step for instance. To complete the tolerance study, 
the variation ranges of the period $\Lambda = L_e + L_f$ and the fill factor $FF = L_f / \Lambda$
 have been evaluated. These extremum values have been plotted on a 2D map in Figure ~\ref{fig_tolerance2}.
 Thus, the variation range of $L_e$, $L_f$, the period $\Lambda$ and the fill factor $FF = L_f / \Lambda$
indicates the area of the $(L_e, L_f)$ couples where the previously defined VCSEL requirements are met. The optimum point (829,~522) with the 
largest bandwidth found by the optimization algorithm is located at the edge of the tolerance area. 
So, in order to enhance the robustness of the grating with respect to the error of fabrication which could be made on $L_e$ and $L_f$,
 a non optimum point located at the centre of the tolerance area has been chosen with $L_e = $ 784 nm and $L_f = $ 537 nm.

The resulting HCG, which characteristics are summarized in Table~\ref{table2}, exhibits a 186~nm large bandwidth
which is 32 nm less than the optimum design reported in Table~\ref{table1}, but is more robust with tolerance values of $\pm$4~\% for $L_e$, $\pm$5~\% for $L_f$
and $\pm$2~\% for $T_g$. These values are not as high as the Si/SiO$_2$-based HCG performances which can exhibit 250~nm large bandwidth
 with up to 10~\% of tolerance. These better performances are due to the higher index contrast of the Si/SiO$_2$ system ($n_{SiO2} = $ 1.509).

\section{Conclusion}
In this work, a Si/Si$_3$N$_4$ high contrast grating mirror has been designed and exhibit a 186~nm large 99.9~\% high reflectivity bandwidth.
The dimension of the mirror structure has been selected by an optimization algorithm which maximizes a quality factor specially defined for a VCSEL application.
 Then, the tolerances of this optimum design have been evaluated and the robustness with respect to the error of fabrication have been improved
by adjusting the grating parameters. This mirror, with a thickness of less than \unit{2}{\micro\meter},
 grooves larger than \unit{500}{\nano\meter} and several percent of tolerance on the grating parameter, should allow laser
operation of VCSEL devices at \unit{2.65}{\micro\meter}.

\section*{Acknowledgment}

The authors thank the French ANR for financial support in the framework of Marsupilami project (ANR-09-BLAN-0166-03) and IES and LAAS (France), partners of LMOPS/Sup\'elec in this project. This work was also  partly funded by the InterCell grant (http://intercell.metz.supelec.fr) by INRIA and R\'egion Lorraine (CPER2007).


\bibliographystyle{spphys}       


\end{document}